\documentclass{aastex}     

\begin{document}

\newcommand{\etal} {{\it et al.}}
\newcommand{\hi} {\ion{H}{1}}
\newcommand{\dg} {$^\circ$}
\newcommand{\uG} {$\mu$Gauss}
\newcommand{\uJy}{$\mu$Jy}
\newcommand{\kps}{km s$^{-1}$}
\newcommand{\pcm}{cm$^{-2}$}
\newcommand{\mjb}{mJy beam$^{-1}$}


\title{The Structure of the Cold Neutral ISM on 10-100 Astronomical
Unit Scales}

\author{M. D. Faison\altaffilmark{1, 2, 3} and W. M. Goss\altaffilmark{2}}

\altaffiltext{1}{Department of Physics and Astronomy, Northwestern University,
Evanston, IL 60208}
\altaffiltext{2}{National Radio Astronomy Observatory, PO Box O,
Socorro, NM 87801}
\altaffiltext{3}{University of Wisconsin, Madison, Dept. of Astronomy,
Madison, WI  53706}

\authoraddr{Address correspondence regarding this manuscript to:
		Michael Faison, Dept. of Physics and Astronomy, 2131
		Sheridan Rd., Evanston, IL 60208}

\begin{abstract}
We have used the Very Long Baseline Array (VLBA) and the Very Large
Array (VLA) to image Galactic neutral hydrogen in absorption towards
four compact extragalactic radio sources with 10 milliarcsecond
resolution.  Previous VLBA data by Faison et al (1998) have shown the
existence of prominent structures in the direction of the
extragalactic source 3C~138 with scale sizes of 10-20 AU with changes
in HI optical depth in excess of 0.8 $\pm$ 0.1.  In this paper we
confirm the small scale \hi\ optical depth variations toward 3C~147
suggested earlier at a level up to 20 \% $\pm$ 5\% .  The sources
3C~119, 2352+495 and 0831+557 show no significant change in \hi\
optical depth across the sources with one sigma limits of 30\%, 50\%,
and 100\%.  Of the seven sources recently investigated with the VLBA
and VLA , only 3C~138 and 3C~147 show statistically significant
variations in HI opacities.

Deshpande (2000) have attempted to explain the observed small-scale
structure as an extension of the observed power spectrum of structure
on parsec size scales.  The predictions of Deshpande (2000) are
consistent with the VLBA \hi\ data observed in the directions of a
number of sources, including 3C~147, but are not consistent with our
previous observations of the \hi\ opacity structure toward 3C~138.

\end{abstract}

 \section{Introduction}
Structure in the interstellar medium (ISM) is observed on many
different size scales, from large kiloparsec scale superbubbles
(Dewdney and Lozinskaya 1994) to electron density fluctuations in the
ionized neutral medium only a few hundred kilometers in size
(Armstrong, Rickett, and Spangler 1995).  Previous studies have
attempted to characterize the structure of the neutral component of
the ISM by observing neutral hydrogen in emission over a range of
angular scales (e.g. Green 1993; Crovisier and Dickey 1983).  These
studies have been limited by the angular resolution of single dish
telescopes and connected-element interferometers to size scales from
about 0.1 parsec to several hundred parsecs.  The results of these
studies suggested that the neutral ISM is smooth on scales smaller
than about 0.1 pc.  Due to the typically low brightness temperature of
\hi\ in emission on small scales, \hi\ 21 cm emission can not be
detected at milliarcsecond scales using Very Long Baseline
Interferometry (VLBI) techniques.  However, it is possible to image
Galactic \hi\ in absorption towards bright extragalactic continuum
sources, and thereby study the structure of the neutral ISM on angular
size scales from the resolution of the VLBI array (about 5 mas at 21
cm for the VLBA) up to the angular size of the continuum source
($\propto$ 50 to 200 mas for the sources in this paper).  At typical
distances to the absorbing gas of 500 to 1000 pc, these angular scales
correspond to tens to hundreds of Astronomical Units.

The first VLBI observations of very small scale structure in the
neutral ISM were published by Dieter \etal\ (1976).  Observing the
source 3C147 with a single baseline from Hat Creek to the Owens Valley
Radio Observatory (150 km), they observed variations in the ratio of
the \hi\ absorption line visibility amplitude to the 21 cm continuum
visibility amplitude as a function of projected baseline.  This change
is caused by non-uniform coverage of the \hi\ absorption across the
face of the source (Radhakrishnan \etal\ 1972).  Based on the galactic
latitude of the source and a model of the continuum emission
components, they inferred that there was structure in the absorbing
\hi\ with a physical size of 70 AU.  These results, based on modeling
of sparse visibility data, were qualitative, but suggested that there
is structure on very small scales in the diffuse neutral ISM.

Subsequent VLBI observations of AU scale \hi\ structure were published
by Diamond \etal\ (1989).  They observed the three extragalactic radio
continuum sources 3C~147, 3C~138 and 3C~380 using three large
telescopes of the European VLBI Network.  The visibility data from the
observations were not sufficient to image the sources and the \hi\
absorption distribution.  However, the ratio of the line to continuum
visibility amplitudes confirmed the non-uniform \hi\ absorption
towards 3C~147 observed by Dieter \etal\ (1976). Large variations
towards 3C~138 and moderate variations towards 3C~380 were discovered.
Diamond \etal\ inferred structure in the \hi\ gas of 25 AU with
densities of $5 - 10 \times 10^4$ \pcm, assuming that the line of
sight dimension is equal to the measured transverse dimension.  These
conclusions were based on modeling the visibility data rather than
directly imaging the \hi\ optical depth structure.

Confirmation of AU scale structure in Galactic \hi\ independent of the
VLBI data was published by Frail \etal\ (1994).  They observed average
variations in the total \hi\ column density of 10-15\% towards nine
pulsars at three observing epochs over a period of 1.7 years.  Each
pulsar in the Frail \etal\ study probed a range of physical size
scales transverse to the line of sight from 5 to 100 AU. The fact that
column density variations over time were observed towards all of the
pulsars suggests that AU scale structure is common throughout the
ISM. Also, the fact that Frail \etal\ observed changes in the total
\hi\ column density but not in the overall line shape suggests that
the AU scale structure is kinematicly related to the diffuse cold
neutral medium as a whole.

The first images of AU scale \hi\ structure in absorption were
published by Davis, Diamond, and Goss (1996) (Henceforth DDG).  They
used the MERLIN array with five EVN telescopes to image 3C~147 and
3C~138 in \hi\ optical depth at an angular resolution of 50 $-$ 150
mas.  They observed significant optical depth variations in one of the
two
\hi\ absorption lines towards both objects.  The largest optical depth
variation of $\Delta\tau_{HI} = 0.16 \pm 0.03$ across the source was
observed in the 1.3 \kps\ absorption line towards 3C~138.  A prominent
optical depth change of $\Delta\tau_{HI} = 0.09 \pm 0.01$ was also
observed in the -8 \kps\ absorption line towards 3C~147.

There is also evidence for AU scale structure in the diffuse molecular
gas.  Moore and Marscher (1995; as well as Marscher \etal\ 1993) have
discussed changes observed in the column density of H$_2$CO towards
the compact objects 3C~111, BL~Lac, and NRAO~150.  They observed the
three sources over 3.4 years with the VLA in the 6 cm H$_2$CO line.
The motion of the sources due to parallax as well as the proper
motions of the absorbing gas caused the relative line of sight through
the molecular gas to the extragalactic sources to change with time.
They observed significant variations in the H$_2$CO opacities towards
3C~111 and NRAO~150, and inferred structure on linear scales $\sim$ 10
AU and H$_2$ densities $\sim 10^6$ cm$^{-3}$, sizes and densities
similar to those observed for the small scale \hi\ structure.

Heiles (1997) has proposed a geometrical model to explain the large
variations in column density.  The geometrical model assumes that the
structures are due to filaments or sheets observed edge-on, causing
larger variations in column densities over smaller scales to be
observed and implying much higher volume densities than are actually
present.  If the observed structure arises from edge-on sheets, the
optical depth images should statistically show smaller-scale
variations along the axis perpendicular to the sheet compared to
variations along the axis parallel with the sheet.

Faison, Goss, Diamond, and Taylor (1998), hereafter FGDT, published
high angular resolution (10-20 mas) VLBA images of \hi\ absorption
towards 3C~138, 2255+416, and 0404+768.  Pronounced \hi\ optical depth
variations were confirmed towards 3C~138, consistent with Diamond
\etal\ (1989) and DDG.  Moderate \hi\ variations were observed towards
the small angular size source 2255+417, and no significant variations
were observed towards 0404+768, suggesting a minimum size of the \hi\
structure of a few tens of AU.

Deshpande, Dwarakanath, and Goss (2000) have evaluated the structure
observed in \hi\ absorption towards Cas A over a linear scale of 4pc
to 0.02 pc. Deshpande (2000) has taken the slope of the power spectrum
of fluctuations of $\alpha = -2.75$ and extrapolated to scales smaller
by factors of three orders of magnitude (sizes scales smaller than 100
AU) and predicts that optical depth variations as large as 0.1 should
be expected on scales of tens of AU. The cause of these fluctuations
is due to gradients in the distribution caused by the ``red'' spectrum
(larger scales have more power) of the power spectrum of the HI. The
fluctuations observed by the VLBA in the direction of 3C~138 are
almost an order of magnitude larger than this prediction. In addition,
two dimensional HI features are observed.

In this paper we present \hi\ optical depth images towards 3C~147,
3C~119, 2352+495, and 0831+557, and we discuss our current knowledge
of AU scale structure in the diffuse neutral ISM.

\section{Observations and Data Reduction}
FGDT discuss the criteria for selecting sources for imaging in \hi\
absorption with the VLBA. The continuum source must be bright enough
at 21 cm to be imaged with acceptable signal-to-noise at sufficient
angular and velocity resolution (continuum flux density greater than
about 100 \mjb\ at 5-10 mas resolution).  The source must have
significant milliarcsecond structure that can be resolved by the VLBA
at 21 cm.  The source must have significant Galactic \hi\ absorption
but not be optically thick ($0.1 < \tau_{HI} < 2.0$).  The angular
resolution of the VLBA at 21 cm is approximately 5 milliarcseconds
(mas).  To improve the signal-to-noise ratio in the optical depth
cube, we have smoothed the images to 10 mas angular resolution.  The
source 3C~147 was selected based on previous observations of \hi\
small scale structure.  The source 3C~119 was selected as a good
candidate for probing small scale \hi\ based on the \hi\ absorption
spectrum towards the source published by Mebold \etal\ (1981) and the
VLBI continuum images published by Nan \etal\ (1991).  The two sources
0831+557 and 2352+495 were selected from the Caltech-Jodrell Bank
survey (Polatidis \etal\ 1995) based on the VLBI angular structure and
the low Galactic latitude of the sources.

The basic data reduction procedure is described in FGDT. Each source
was observed with all of the ten VLBA stations and the phased VLA (as
an additional VLBI station) for 12 hours.  The data were correlated at
the VLBA correlator in Socorro, New Mexico.  Four separate IFs were
used, with one IF centered on the absorption lines and three IFs
separated by at least 100 \kps\ in velocity to sample the 21 cm
continuum emission.  For all of these observations the data were
correlated with velocity channels of 0.4 \kps, and for most of the
objects a 500 kHz bandwidth per IF was used with 256 spectral channels
per IF.  For the observations of 3C~119 a bandwidth of 1 MHz per IF
was used with 512 spectral channels per IF to accommodate the larger
velocity range (100 \kps) of \hi\ absorption components.

All data processing for each object was done using the AIPS
(Astronomical Image Processing System) package available from
NRAO. The data were amplitude calibrated and fringe fitted using
standard techniques in AIPS.  Once the data were amplitude and
bandpass calibrated, the three continuum IFs were averaged together,
self-calibrated, and imaged using the AIPS task IMAGR. The
self-calibration phase and amplitude solutions from these continuum
images were then applied to the line IF, which was then imaged to
produce a total intensity cube as a function of RA, Dec, and LSR
velocity.  The velocity channels in this intensity cube that did not
show \hi\ absorption were then averaged together to produce a
continuum image.  This image was used with the intensity cube to
produce an \hi\ optical depth cube as a function of RA, Dec, and LSR
velocity.  The optical depth cube is calculated as $\tau_{HI} (RA,
Dec, V) = - \ln [{I_{line} (RA, Dec, V) / I_{continuum} (RA, Dec)}]$.
Because the signal-to-noise ratio in the optical depth images is poor
where the continuum emission is weak, the optical depth images were
blanked where the continuum emission was less than 10\% of the peak
emission.  This optical depth cube with dimensions of RA, Dec, and
velocity is the final reduced data for each object.

\section{Results}

\subsection{3C~147}
\label{3C137sec}

The continuum image of 3C~147 is shown in Figure~\ref{3c147cont}.
This image was convolved to a resolution of 10 mas.  The source is a
quasar with a redshift of 0.545 and a total 21 cm flux density of 22.5
Jy.  The continuum structure consists of a core component with a broad
jet extended to the SW.  Figure~\ref{3c147tauspec} shows an \hi\
optical depth spectrum averaged over the entire source.  Kalberla,
Schwarz, and Goss (1985) deconvolved the \hi\ optical depth spectrum
towards 3C~147 into five Gaussian components.  In this paper we will
only consider the three components at 0.4, $-$8.3, and $-$10.3 \kps.
Figures~\ref{3c147taumap119} and \ref{3c147taumap140} show \hi\
optical depth images towards 3C~147 at 0.4 and $-$8.3 \kps,
respectively.  The greyscale indicates \hi\ optical depth from 0 to
1.5 and the contours indicate the low-level continuum emission.
Figure~\ref{3c147fits} shows the 21 cm continuum image with \hi\
optical depth spectra averaged over a beam size of 10 mas towards the
three areas of brightest emission (components A, B, and C in
Figure~\ref{3c147cont}).  The dashed line in these plots indicates the
averaged spectrum, and the solid line indicates the Gaussian fit
produced using the iterative fitting task PROFIT, using the Groningen
Image Processing System (GIPSY) package .  The parameters of these
Gaussian profile fits are given in Table~\ref{3c147comp}.

In the optical depth image at $-$8.3 \kps\
(Figure~\ref{3c147taumap140}), there is a increase in optical depth
from 0.82 to 1.1( $1 \sigma$ is 0.05) between the bright NE component
to the SW component 190 mas away, consistent with that found by DDG.
They observed \hi\ optical depth variations in the $-$8.3 \kps\
absorption line towards 3C~147, but not in the 0.4 \kps\ absorption
feature.

In the 0.4 \kps\ optical depth image (Figure~\ref{3c147taumap119}),
there is a slight change in optical depth between the two brightest
components.  The largest optical depth change at this velocity is from
$\tau = 0.5 \pm 0.1$ to $0.7 \pm 0.1$ between the bright NE component
and the component midway along the axis of the source.  These are
small angular scale optical depth variations that would not have been
resolved at the 150 mas angular resolution of the MERLIN images
published by DDG.

A problem with determining the physical size of this structure is the
uncertain determination of the distance to the absorbing \hi\ gas.
Davis, Diamond, and Goss (1996) placed a simple geometrical upper
limit on the distance to the \hi\ cloud based on the Galactic latitude
of the source with $b=+10.3$\dg.  Assuming that the cold \hi\ is
confined to 100 pc of the plane, an intervening \hi\ cloud along this
line of sight must be within 600 pc of the sun.  At a Galactic
longitude of $l=162$\dg, a kinematic distance to the $-$8.3 \kps\
component is uncertain; however, an estimate using the rotation curve
of Fich, Blitz, and Stark (1989) for the distance to the $-$8.3 \kps\
is 1.1 kpc.  The angular separation between components A and C of 190
mas corresponds to a physical scale of 200 AU at a distance of 1 kpc
and 100 AU at 600 pc.  The angular separation between components A and
B of 83 mas corresponds to 90 AU at 1 kpc and 50 AU at 600 pc.  These
distances are quite uncertain, and so it is reasonable to approximate
the distance to the -8.3 \kps\ gas at 1kpc and the distance to the low
velocity gas at 500 pc.

As with the observations described by FGDT, the circular polarization
($V = RCP - LCP$) was imaged for each source in addition to the total
intensity cube.  The circular polarization cube can be fit for the
Zeeman splitting due to a line of sight magnetic field.  We fit the
circular polarization spectrum of these sources using the tasks
ZEESTAT and ZEESIM in the MIRIAD data analysis package described by
Sault \etal\ (1995).  These tasks average the $I$ and $V$ spectra over
the entire source and estimate the leakage of the $I$ spectrum in the
$V$ spectrum using an iterative fitting procedure.  The strength of
the field can be calculated from the $V$ and $I$ spectra as $B_{||} =
{2 \, V \over a\, dI/d\nu}$, where $a = 2.8$ Hz \uG$^{-1}$.  For
3C~147, the fit gives a $2\sigma$ upper limit on the absolute line of
sight magnetic field averaged over the entire source of 32 \uG.

\subsection{3C~119}
\label{3C119sec}
3C~119 is a compact symmetric object with a redshift of 0.408 (Nan
\etal\ 1999)and  total flux density of 8.6 Jy.
Previous observations of the Galactic \hi\ absorption spectrum towards
3C~119 by Mebold \etal\ (1981) indicated significant \hi\ absorption
along the line of sight. The 21 cm continuum image is shown in
Figure~\ref{3c119cont}, convolved to 10 mas angular resolution.  Only
three bright continuum components are observed in the image, labeled
``A'', ``B'', and ``C'' in Figure~\ref{3c119cont}.

The source 3C~119 has the lowest galactic latitude ($b=-4.3$\dg) of of
the seven sources in the Galactic \hi\ sample.  The source has a
complex \hi\ absorption spectrum with Galactic \hi\ absorption over a
velocity range from $+10$ to $-63$ \kps\.  The \hi\ optical depth
spectrum towards 3C~119 is shown in Figure~\ref{3c119tauspec}.  Two
typical optical depth images for 3C~119 are shown in
Figures~\ref{3c119taumap187} and \ref{3c119taumap327} for LSR
velocities of $-$4.7 \kps\ and $-$62.6 \kps, respectively.  The inset
plots show crosscuts from south to north in optical depth through the
centers of the two brightest components ``C'' and ``B''.  The optical
depth images at most velocities show no \hi\ optical depth variations
between the two components B and C ; the $1\sigma$ upper limit to any
change in the opacity is $\Delta\tau \approx 0.2$.

Figure~\ref{3c119fits} shows the 21 cm continuum image with optical
depth spectra towards continuum components B and C, averaged over a
beam size of 10 mas.  The angular separation of these components is
about 53 mas.  These two average spectra were fit with eight Gaussian
profiles using PROFIT in the GIPSY data reduction package. The
parameters of these two sets of fits are given in
Table~\ref{3c119comp}.  As with Figure~\ref{3c147fits} for 3C~147, the
dashed lines indicate the data and the solid lines indicate the fits.
There is no significant difference in any of the profile fits between
the continuum emission components B and C with a $1 \sigma$ upper
limit of about 10\% in column density variation.

The source 3C~119 is at a Galactic longitude of $l=161$\dg, and a
kinematic distance to the \hi\ absorption components can not be
reliably determined this close to the Galactic anti-center.  A
geometric upper limit on the distance to the \hi\ absorption given the
galactic latitude of the source ($b=-4.3$\dg) is 1.3 kpc, again
assuming that the \hi\ layer is within 100 pc from the plane.  At that
distance, the angular resolution in the optical depth images
corresponds to a physical size of 13 AU, and the 53 mas separation
between components B and C corresponds to 70 AU.  The low velocity
absorption components are likely due to local gas (closer than 500
pc), and the \hi\ gas at high negative velocity is probably in the
Perseus arm at a distance greater than 2 kpc.  At that distance, the
10 mas beam in the \hi\ optical images corresponds to 20 AU, and the
separation of 53 mas between components B and C corresponds to 100 AU.

As with 3C~147, we fit the circular polarization spectrum towards
3C~119 to determine the line of sight magnetic field.  A fit to the
overall $V$ and $I$ spectra averaged over the entire source produces a
$2 \sigma$ upper limit on the line of sight magnetic field of $20$
\uG.

\subsection{2352+495}
\label{2352sec}
The source 2352+416 is a compact symmetric object selected from the
Polatidis \etal\ (1995) VLBI survey.  There are no published previous
observations of \hi\ absorption towards this source , a radio galaxy
with a redshift of 0.237, and a total 21 cm continuum flux density of
2.36 Jy.  The morphology of the source is a bright core with two
symmetric lobes.  The VLBA 21 cm continuum image of 2352+495 is shown
in Figure~\ref{2352cont}, with the three bright continuum components
from north to south labeled ``A'', ``B'', and ``C''. The angular
resolution is 10 mas.

The \hi\ optical depth spectrum towards 2352+495 is shown in
Figure~\ref{2352tauspec}.  The maximum \hi\ optical depth is 0.25 and
occurs at a velocity of 3.3 \kps\ with a line width of 2.0 \kps.  The
\hi\ absorption line is asymmetrical and consists of several narrow
components, with the strongest line at 0 \kps.
Figure~\ref{2352taumap112} shows the optical depth image at the
velocity of the maximum \hi\ optical depth at v=0 \kps.

Figure~\ref{2352fits} shows the continuum image with average \hi\
optical depth spectra towards the three bright emission components.
As with Figures~\ref{3c147fits} and \ref{3c119fits}, the dashed lines
indicate the VLBA data and the solid lines indicate the Gaussian
profile fits.  The parameters of these Gaussian profile fits are given
in Table~\ref{2352comp}. Three Gaussian profiles were fit to the
spectra; however, spectral components 2 and 3 (indicated in
Figure~\ref{2352tauspec}) are considerably noisier than spectral
component 1, and the Gaussian profile fits for those components are
not well-determined.  These data are consistent with uniform \hi\
absorption across the source with no changes in \hi\ column density
exceeding 30\% ($1\sigma$).

The source 2352+495 is located $l=113$\dg\ and $b=-12.0$\dg.  The two
spectral components near 0 \kps\ are assumed to be local gas.
Assuming that the thickness of the local \hi\ disk is less than 100
pc, \hi\ gas along a line of sight at this Galactic latitude must lie
closer than 500 pc.  The resolution of the \hi\ optical depth images
of 10 mas corresponds to a linear size of 5 AU, and the separation of
50 mas between continuum components A and C corresponds to 25 AU.
Assuming the rotation curve of Fich, Blitz, and Stark (1989), the
kinematic distance to the $-$10 \kps\ absorption line is 800 pc.  At
this distance, the 10 mas beam corresponds to a linear size of 8 AU,
and the 50 mas separation between the components A and C corresponds
to 40 AU.

Fits to the $V$ and $I$ spectra averaged over the entire source
produce a $2\sigma$ upper limit on the line of sight magnetic field of
$160$ \uG.

\subsection{0831+557}

The source 0831+557 was also selected from the VLBI survey of
Polatidis \etal\ (1995). There are no published previous observations
of \hi\ absorption towards this source.  This source has a total 21 cm
continuum flux density of 8 Jy.  It has the highest Galactic latitude
of any of the sources in the sample with $b=34.6$\dg.  It has very
weak \hi\ absorption in two velocity components, at 3 \kps\ with a
maximum \hi\ optical depth $\approx 0.1$, and a possible weak line
near $-$37 \kps\ with a maximum \hi\ optical depth of $\approx 0.03$.
The 3
\kps\ line has a velocity width of 2 \kps.  Figure~\ref{0831cont}
shows a 21 cm continuum image of 0831+557, convolved to an angular
resolution of 20 mas.  There are two strong continuum components
separated by about 180 mas, labeled ``A'' and ``B'' in
Figure~\ref{0831cont}.  Component A is significantly brighter, with a
peak continuum emission of 2.9 Jy beam$^{-1}$, compared to 0.4 Jy
beam$^{-1}$ peak emission for Component B.

Figure~\ref{0831taumap3} shows an optical depth image towards 0831+557
at 3.9 \kps.  The greyscale indicates \hi\ optical depth from 0 to
0.2, and the contours represent weak continuum emission.  The image
was convolved to 20 mas to improve the signal-to-noise ratio.  The
inset plot shows a cross cut in optical depth along the dashed line.
No significant variations in optical depth are observed between the
two continuum components with a $1 \sigma$ limit in any variation of
$\Delta\tau \approx 0.1$.

At a galactic latitude of $b=+36.6$\dg, a geometric upper limit to the
absorbing gas is $\propto 200$ pc, assuming the thickness of the cold
\hi\ gas is 100 pc.  At that distance, the angular separation of 160
mas between components A and B corresponds to a physical distance of
30 AU.

\section{Structure Functions of the \hi\ optical depth images}

Deshpande (2000) has pointed out that the optical depth variations
observed between two points on the sky represent contributions in
optical depth variations on all scales, including larger scales.  The
optical depth difference between two points on the sky do not arise
entirely from structures with this scale size.  Due to the shape of
the power spectrum (a ``red`` spectrum with enhanced fluctuations at
larger scales), variations from larger structures contribute to the
variations at an observed scale.  This effect complicates the
interpretation of the optical depth images as representing unique
structure on tens of AU scales.  A proper interpretation of the actual
physical scale of the structure requires knowledge of the power
spectrum over larger scales.  Deshpande, Dwarakanath, and Goss (2000)
have found that the power spectrum of \hi\ optical depth variations
towards Cas A (Perseus arm feature) and the Outer Arm towards Cyg A
can be fit with a power law index of $-2.75$ over physical scales from
0.02 pc to 4 pc.  The Local Arm in the direction of Cyg A, the slope
is shallower at $-$2.5.  The power spectra of both Cyg A features are,
however, an order of magnitude less intense than the Cas A Perseus Arm
feature.  Deshpande (2000) suggests that this power law power spectrum
with a steeper index of $-2.75$ can be extrapolated to size scales of
40$--$100 AU, a factor of almost 40 relative to the smallest scale in
the measured Cas A power spectrum. Thus optical depth variations would
be expected with root-mean-square values of about 0.1.  This effect is
shown in Figure 2 from Deshpande (2000), based on the -2.75 index for
the Cas A Perseus arm data.  Most of the objects in this paper and
FGDT show \hi\ optical changes which are consistent with this level of
variation.  However, 3C~138 shows changes in excess of this
prediction.  For 3C~138 the changes in opacity are in the range
0.6$-$0.8 over a size scale of 50 mas (25 AU at a distance of 500 pc).

Deshpande (2000) also suggests that structures observed on small
scales should only be observed as gradients in optical depth, since
the changes on AU scales are only due to larger scale variations.  Our
\hi\ optical images toward 3C~147 and 3C~138 clearly show discrete
\hi\ structures, not simply gradients.  Especially convincing is the
elongated two dimensional structure in the NE part of the image where
the continuum emission of 3C~138 is very strong, and therefore the
\hi\ optical depth is well determined.  This part of the \hi\ optical
depth image at 0.4 \kps\ from FGDT is reproduced in
Figure~\ref{3c138round}.  These stuctures clearly are {\i not} simply
gradients in optical depth.

\section{Discussion and Conclusions}

The optical depth variations towards 3C~147 are shown in the
position-velocity plot for 3C~147, shown in Figure~\ref{3c147lv}.  The
vertical axis of this position-velocity corresponds to the axis along
the source from the NW to the SE as indicated in
Figures~\ref{3c147taumap119} and \ref{3c147taumap140}.  As can be
noticed in the individual optical depth images, the optical depth
variations along the axis of the source are larger in \hi\ absorption
line near $-8.3$ than in the line near $0.4$ \kps.  The fluctuations
in the opacity in Figure~\ref{3c147lv} far away from the absorption
lines are a good indication of the noise in the optical depth at a
given position along the axis of the source.  This position-velocity
plot can be compared with the position-velocity plot for 3C~138 from
FGDT.  Although there are obvious optical depth variations in each
absorption line towards 3C~147, the striking optical depth structure
changes in velocity observed towards 3C~138 are not observed in the
3C~147 position-velocity image.  In the 3C~138 position-velocity plot,
the velocity features that slope diagonally across are quite striking
and indicate that we are observing real, three dimensional features
coherent in velocity.

The constraints on the line of sight magnetic field towards these
sources suggest that the magnetic field towards these sources is not
enhanced relative to the average interstellar magnetic field of $40 -
100$ \uG .  Diamond \etal\ (1989) determined that field strengths of
200 \uG\ would be required to confine the small scale structure
observed towards 3C~138 relative to the pressure of the diffuse ISM.
The upper limits on the line of sight magnetic field towards 3C~147
and 3C~138 are an order of magnitude below this value.

In conclusion, we have imaged Galactic \hi\ in absorption towards four
extragalactic continuum objects at 10 milliarcsecond resolution.  The
-8.3 \kps\ \hi\ absorption line towards the source 3C~147 shows
variations of $10$ to $20$ \% in \hi\ column density across the source
($1 \sigma$ error of 5\%), and variations in optical depth from about
0.82 to 1.1 (at about a $4 \sigma$ level).  No significant variations
\hi\ optical depth are observed across the source 3C~119 to within a
$1 \sigma$ error of 0.2 in optical depth and a 10\% change in \hi\
column density.  The source 2352+495 shows no significant \hi\ column
density variations in column density to within a $1 \sigma$ limit of
30\%.  The high Galactic latitude source 0831+557 has weak \hi\
absorption and shows no significant column density variations to
within a $1 \sigma$ limit of 50\%.

All of these results are consistent with the prediction of Deshpande
(2000) that optical depth variations as large as $\delta\tau \approx
0.1$ on scales of 100's of AU.  However, the source 3C~138 is unique
among the objects we have observed in showing very large \hi\ optical
depth variations of $\Delta\tau \approx 0.1$ on scales of 10's of AU.

It is of considerable interest to probe the cold, neutral medium on
intermediate scales between the few tens of AU probed by VLBI
observations and the 0.1 pc scales probed by connected-element
interferometers such as the WSRT and the VLA.  The only observations
published to date on these angular scales between the resolution of
the VLBA at about $10-500$ mas and the VLA at about $1-30$ \arcsec\
are observations of ISM optical lines towards binary stars (e.g.,
Lauroesch \etal\ 1999; Watson and Meyer 1996) separated by 500 to
30,000 AU.  However, most of these observations are of variations in
\ion{Na}{1} column density, which may not reliably trace the \hi\
column density.  Observations of \hi\ column density variations should
be carried out on these size scales to compare the AU scale structure
with the larger scale structure.  These observations could be carried
out with the MERLIN array and also by observing \hi\ column density
variations towards pulsars as done by Frail \etal\ (1994).  The Green
Bank Telescope will be an especially sensitive instrument for
observing \hi\ in absorption towards pulsars.  Also, more pulsar and
VLBI studies of AU scale structure should be carried out in the
southern hemisphere, outside the field view of the VLBA.  To date no
observations of AU scale structure in \hi\ have been made from
$l=187$\dg\ to $l=20$\dg.

\acknowledgements 

The VLBA and VLA are operated by the National Radio Astronomy
Observatory, a facility of the National Science Foundation, operated
under a cooperative agreement by Associated Universities, Inc.  We
thank Ed Churchwell for helpful comments on an earlier version of this
paper.  We also thank A. Deshpande for providing the structure
function code and assisting us with interpretation of the structure
functions of the optical depth images.


\begin{table} 
\dummytable\label{srcparams}
\end{table} 

\begin{table}
\dummytable\label{3c147comp}
\end{table}

\begin{table}
\dummytable\label{3c119comp}
\end{table}

\begin{table}
\dummytable\label{2352comp}
\end{table}

\newpage

\figcaption[Faison.fig1.ps]{Continuum image of 3C~147 at 21 cm
obtained with the VLBA and phased VLA.  The beam has been convolved to
an angular size of 10 milliarcseconds from an original resolution of
$5\times 4$ mas.  The contours correspond to the continuum emission at
23.1, 46.2, 115, 231, 462, 694, 924, and 1160 \mjb. \label{3c147cont}}

\figcaption[Faison.fig2.ps]{\hi\ optical depth spectrum towards 3C~147
averaged over the entire source using the VLBA and phased VLA.  The
velocity resolution is 0.4 \kps. \label{3c147tauspec}}

\figcaption[Faison.fig3.ps]{\hi\ optical depth image towards 3C~147 in a single 0.4
\kps\ velocity channel at 0.4 \kps\ LSR velocity.  The angular
resolution is 10 mas.  The greyscale represents \hi\ optical depth
from 0 to 1.0, and the contours represent 21 cm continuum emission at
23.1, 46.2, 116 \mjb.  The optical depth image has been blanked where
the continuum emission is weak (where the error in the optical depth
is greater than 0.1).  The inset plot shows a cross cut in optical
depth with 2 $\sigma$ errorbars along the axis indicated by the dashed
line. \label{3c147taumap119}}

\figcaption[Faison.fig4.ps]{\hi\ optical depth image towards 3C~147 in a single 0.4
\kps\ velocity channel at $-$8.3 \kps\ (LSR).  The angular resolution
is 10 mas.  The greyscale represents \hi\ optical depth from 0 to 1.3,
and the contours represent 21 cm continuum emission at 23.1, 46.2, 116
\mjb.  The optical depth image has been blanked where the continuum
emission is weak (where the error in the optical depth is greater than
0.1).  The inset plot shows a cross cut with 2 $\sigma$ errorbars in
optical depth along the axis indicated by the dashed line.
\label{3c147taumap140}}

\figcaption[Faison.fig5.ps]{Continuum image of 3C~147 with inset plots showing
optical depth spectra averaged over the areas of brightest continuum
emission.  The dashed line in the inset plots shows the VLBA data, and
the solid line indicates the Gaussian component fits.  The parameters
of the fits are given in Table~\ref{3c147comp}. \label{3c147fits}}

\figcaption[Faison.fig6.ps]{Continuum image of 3C~119 at 21 cm from the VLBA and
phased VLA.  The beam has been convolved to an angular size of 10 mas.
The contours represent continuum emission at 169, 337, 674, 1010,
1350, and 1690 \mjb. \label{3c119cont}}

\figcaption[Faison.fig7.ps]{\hi\ optical depth spectrum towards 3C119,
averaged over the source. The velocity resolution is 0.4 \kps.  The
numbered velocity components are referred to in
Table~\ref{3c119comp}. \label{3c119tauspec}}

\figcaption[Faison.fig8.ps]{\hi\ optical depth image towards 3C~119 in a single 0.4
\kps\ velocity channel centered at $-$4.7 \kps\ (LSR).  The greyscale
represents \hi\ optical depth from 0.35 to 0.85, and the contours
represent continuum emission at 112, 374, and 748 \mjb.  The angular
resolution is 10 mas.  The inset plot shows a cross cut with 2
$\sigma$ errorbars in optical depth along the axis indicated by the
dashed line. \label{3c119taumap187}}

\figcaption[Faison.fig9.ps]{\hi\ optical depth image towards 3C~119 in a single 0.4 \kps\
wide velocity channel centered at $-$62.6 \kps\ LSR velocity.  The
greyscale represents \hi\ optical depth from 0.1 to 0.4, and the
contours represent continuum emission at 112, 374, and 748 \mjb.  The
resolution of the beam is 10 mas.  The inset plot shows a cross cut
with 2 $\sigma$ errorbars in optical depth along the axis indicated by
the dashed line. \label{3c119taumap327}}

\figcaption[Faison.fig10.ps]{Continuum image of 3C~119 with inset plots showing
optical depth spectra averaged over the areas of brightest continuum
emission.  The dashed line in the inset plots shows the VLBA data, and
the solid line indicates the Gaussian component fits.  The parameters
of the fits are given in Table~\ref{3c119comp}. \label{3c119fits}}

\figcaption[Faison.fig11.ps]{Continuum image of 2352+495 at 21 cm from the VLBA and
the phased VLA. The beam has been convolved to 10 mas.  The contours
correspond to continuum emission at 91, 182, 273, 364, 455, 547, 638,
729, and 820 \mjb. \label{2352cont}}

\figcaption[Faison.fig12.ps]{\hi\ optical depth spectrum towards 2352+495, averaged
over the source.  The velocity resolution is 0.4 \kps\ per channel.
\label{2352tauspec}}

\figcaption[Faison.fig13.ps]{\hi\ optical depth image towards 2352+495 in a single
0.4 \kps\ velocity channel centered at 3.3 \kps\ (LSR).  The
resolution of the beam is 10 mas.  The greyscale corresponds to \hi\
optical depth from 0.1 to 0.3, and the contours represent continuum
emission at 18, 46, and 91 \mjb.  The inset plot shows a cross cut
with 2 $\sigma$ errorbars in optical depth along the axis indicated by
the dashed line. \label{2352taumap112}}

\figcaption[Faison.fig14.ps]{Continuum image of 2352+495 with inset plots showing
optical depth spectra averaged over the areas of brightest continuum
emission.  The dashed line in the inset plots shows the VLBA data, and
the solid line indicates the Gaussian component fits.  The parameters
of the fits are given in Table~\ref{2352comp}. \label{2352fits}}

\figcaption[Faison.fig15.ps]{Continuum image of 0831+557 at 21 cm from the VLBA and
the phased VLA. The beam has been convolved to 20 mas.  The contours
correspond to continuum emission at -29.2, 29.2, 58.3, 146, 292, 583,
and 1460 \mjb.  \label{0831cont}}

\figcaption[Faison.fig16.ps]{\hi\ optical depth spectrum towards
0831+557 observed with the VLA.  The velocity resolution is 1.3 \kps.
\label{0831tauspecvla}} 

\figcaption[Faison.fig17.ps]{\hi\ optical depth image towards 0831+557
in a single 0.4 \kps\ velocity channel centered at 3.9 \kps\ (LSR).
The greyscale represents \hi\ optical depth from 0.0 to 0.2, and the
contours represent continuum emission at 29.2, 58.3, and 292 \mjb.
The angular resolution is 20 mas.  The inset plot shows a cross cut
with 2 $\sigma$ errorbars in optical depth along the axis indicated by
the dashed line. \label{0831taumap3}}

\figcaption[Faison.fig18.ps]{A section of the \hi\ optical depth image
at 0.9 \kps\ of 3C~138, taken from FGDT. Two cross sections in optical
depth are shown with 2$\sigma$ errors.
\label{3c138round}}

\figcaption[Faison.fig19.ps]{Position-velocity plot for 3C~147 along
the axis of the source indicated by the dashed line in
Figures~\ref{3c147taumap119} and \ref{3c147taumap140}.  The horizontal
axis indicates LSR velocity, and the vertical axis indicates angular
distance from the upper NW corner of the source.  The greyscale
indicates \hi\ optical depth from 0.0 to 1.0.  Variation in optical
depth can be observed as a change in the greyscale along a vertical
line in this plot.  More prominent variations in optical depth are
observed along the source axis at the $-8.3$ \kps\ line than at the
0.4 \kps\ line . \label{3c147lv}}

\end{document}